\documentclass[amsmath,amssymb,aip,jap]{revtex4-2}
\usepackage{xcolor}
\usepackage{booktabs}
\usepackage{graphicx}
\usepackage{dcolumn}
\usepackage{bm}
\begin{document}
\title{Bilayer artificial spin ice: magnetic force switching and basic thermodynamics}

\author{Fabio S. Nascimento} \email{fabiosantos.ba@gmail.com}
\affiliation{Centro de Forma\c{c}\~ao de Professores, Universidade Federal do Rec\^{o}ncavo da Bahia, 45300-000 - Amargosa - Bahia - Brazil.}
\author{Afranio R. Pereira}\email{apereira@ufv.br}
\affiliation{Departamento de F\'{i}sica, Universidade Federal de Vi\c{c}osa, 36570-900 - Vi\c{c}osa - Minas Gerais - Brazil.}
\author{Winder A. Moura-Melo}\email{w.mouramelo@ufv.br}
\affiliation{Departamento de F\'{i}sica, Universidade Federal de Vi\c{c}osa, 36570-900 - Vi\c{c}osa - Minas Gerais - Brazil.}
\begin{abstract}
We study an artificial spin ice system consisting of two identical layers separated by a height offset $h$. For small separation, the layers are shown to attract each other, provided the whole system is in the ground state. Such an attraction comes about by means of a power-law force that we compare to van der Waals forces. When magnetic monopoles occur in one (or both) layers, the scenario becomes even more interesting and these layers may also repel each other. By tuning parameters like $h$ and monopole distance, switching between attraction and repulsion may be accomplished in a feasible way. Regarding its thermodynamics, the specific heat peak shifts to lower temperature as $h$ increases.
\end{abstract}
\maketitle
%
%
\section{Introduction and Motivation}
Geometrical frustration is an interesting phenomenon which has
received a lot of attention recently\cite{MoessnerToday}. In magnetism, it
arises whenever interactions between magnetic degrees of freedom are incommensurate with the lattice underlying the crystal geometry. Frustration emerges in appealing natural materials\cite{Bramwell,Castelnovo,Balents},
and it can also be created by design\cite{Design}. Indeed,
artificial systems have been built in diverse configurations which allow us to control frustration by experimentally tuning suitable parameters. An important class of such designed systems is provided by artificial spin ice
(ASI) arrangements\cite{Wang2006,MolJAP2009,Morgan2011,MolTriangular,Moler09,Zhang2013,Mol3D,ReviewASIgeometries-Nisoli,Advances2020,Reviews2019,RectangularASI-2,Loreto18,PRResearch2020,Cuneder2019,Stamps2020},
which essentially consists of a planar-type regular array of nanosized elongated
ferromagnetic rods where geometrical frustration takes place at the vertices. By virtue of strong shape anisotropy along the major axis, every nanoisland effectively behaves as an Ising-like dipole. Now, the collective interaction among all these dipoles yields surprising emergent phenomena, such as fractionalization. Actually, above the ground state the most elementary excitations show up as magnetic monopoles, coupled in pairs by energetic strings\cite{MolJAP2009,Morgan2011},  which are flux-carrying magnetized chains. In other words, the original degrees of freedom, the usual magnetic dipoles, have been {\em fractionalized} into isolated monopoles emerging at ASI vertices. Although they had been originally named {\em Dirac
monopoles and strings}, it is more suitable to speak about Nambu monopoles and strings as claimed in Refs. \cite{Silva2013,Marrows2019}, after Nambu picture adaptation of Dirac description to a London-type framework\cite{Nambu74,Volovik}. Such a phenomenon has been observed to occur in distinct ASI lattice geometries including square, rectangular, triangular, and kagome arrangements.\\

\begin{figure}
    \centering
    \includegraphics[width=13 cm]{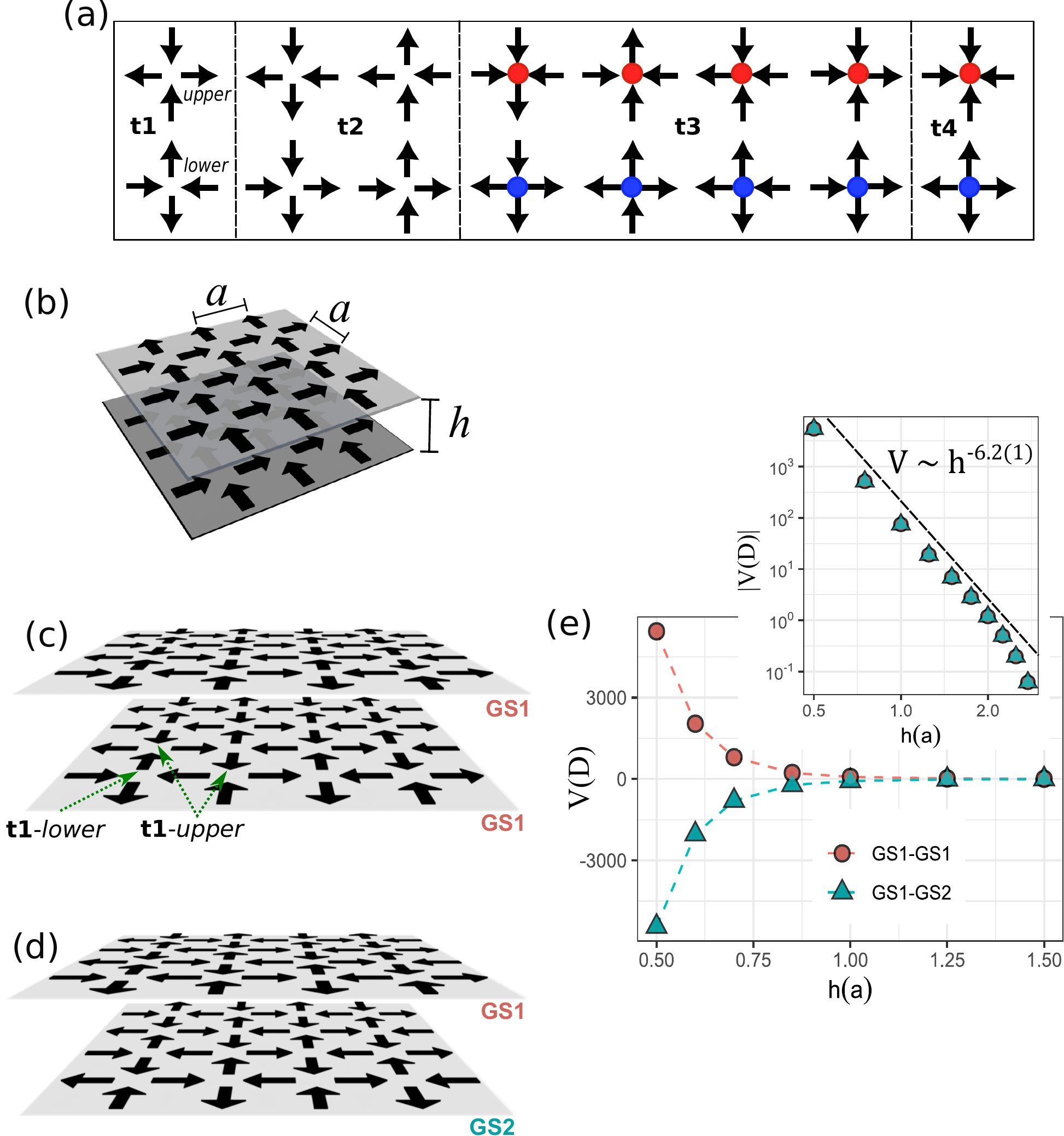}
    \caption {{\bf (a)}: the four possible classes of vertices in a single monolayer square ASI vertex: $t1$, $t2$, $t3$ and $t4$. Each class comprises different vertex types that share the same energy. The first two classes obey the ice-like rule: two spins point {\em in} while the other two point {\em out} of the vertex center, in short {\em 2in-2out}, while the other topologies violate it. {\bf (b)}: sketch of BASI system showing the square arrangement of each layer with lattice spacing $a$ and separated by height offset $h$. ({\bf c}): each layer displays ground state composed only by $t1$ vertices (upper layer is $GS1$). ({\bf d}): now, the bottom layer presents $GS2$. ({\bf e}) potential between the layers is plotted as a function of $h$. It is noteworthy that $GS1-GS2$ is the true combined ground state (for small offset, $h\lesssim  a$). Indeed, in this case, the layers experience a mutual attraction, while for $GS1-GS1$ a repulsive potential tends to keep them apart from each other. Note also that this potential decays rapidly, $V\sim h^{-6.2(1)}$, and practically vanishes as $h\gtrsim 1.5 a$.}
    \label{fig:fig1}
\end{figure}

The interest in their physical properties lies in the fact that such systems are promising candidates for new technologies based upon the control of magnetic charges and their currents, something called magnetricity and magnetronic. Actually, magnetic charge flow was firstly realized in Dy$_2$Ti$_2$O$_7$ compound, an example of natural three-dimensional spin ice crystal, but at very low temperature $\sim 200-300{\rm mK}$, see Refs.\cite{Bramwell2009}. At room temperature, an ordered magnetic current has been observed in an unidirectional arrangement of patterned nanoislands \cite{Loreto2015unidirecional}, where no geometrical frustration takes place at all. In turn, even though theoretical studies regarding three-dimensional ($3{\cal D}$) ASI appeared more than a decade ago \cite{Moler09,Mol3D,Moller2006}, their experimental realization took place only very recently\cite{Rougemaille,Communic2019,Farhan2019}, which has been achieved due to novel advances in patterning $3{\cal D}$ magnetic nanostructures\cite{Fernandez-Pacheco12017}. These systems consist of only one ASI built by offsetting one of the sublattices by some height $h$ such that the energy of interaction between all nearest neighbours becomes equivalent, allowing this arrangement to undergo a transition to a magnetic Coulomb phase \cite{Moler09,Mol3D,Moller2006,Rougemaille}. In order to shed further light onto such phenomena, we investigate a rather similar system whose layers contain nanoislands arranged like a square lattice. More specifically, we study a bilayer artificial spin ice (BASI, for short), where interactions take place among all the islands of both layers (see Fig. \ref{fig:fig1}). Among other results, we realize that layers experience a mutual attraction whenever BASI system is at ground-state. Further, repulsion is brought about when excitations  take place into the system. Now, attraction and repulsion between the layers may be switched by tuning height separation and excitations configurations. Additionally, BASI specific heat peak is shifted to lower temperature as $h$ is increased. The article is outlined as follows: In Section II we present our model and methods, while results and discussion is left to Section III. We close our paper by presenting our conclusions and prospects for forthcoming research.\\
\section{Model and Methods}
The magnetic moment of each square ASI interact via dipolar Hamiltonian, as below:
\begin{equation} \label{Hamiltonian}
\textit{H}_{D}=D a^{3} \sum_{i>j} \left[
\frac{\hat{e}_i\cdot\hat{e}_j}{r^3_{ij}}-\frac{3(\hat{e}_i\cdot\vec{r}_{ij})(\hat{e}_j\cdot\vec{r}_{ij})}{r^5_{ij}}
\right] S_i S_j\,,
\end{equation}
where $D=\frac{\mu_0}{4\pi}\frac{\mu^2}{a^3}$ is the dipole-dipole coupling constant, $\hat{e}_i$ is the local Ising axes of the lattice, $r_{ij}$ is the distance between $S_i$ and $S_j$, and $S_i=\pm 1$ accounts for the unit magnetic moment states which behave as effective Ising spins due to shape anisotropy of elongated nanoislands. [For typical ASI arrangements, each nanoisland carry magnetic moment strength $\mu \sim 10^6 -10^7 \mu_B$ ($\mu_B$ is the Bohr magneton) and they are separated by lattice spacing $a\sim 10^2 \,{\rm nm}$, so that $D\sim 10^{-18} - 10^{-20}\,{\rm J}$]. If one has a single ASI layer, then such vectors run over directions $x$ and $y$, confined on the layer plane. Figure \ref{fig:fig1}a shows the four possible vertex types in a single ASI (obtained from Hamiltonian (\ref{Hamiltonian}) when only intra plane interactions are taken into account). In this case, the ground state is manifested when all vertices are in configurations of class $t1$, obeying the ice rule. Class $t2$ also presents vertices which obey the ice rule but they have larger energy than vertices belonging to $t1$. In general, vertices type $t2$ are associated with the strings connecting monopoles living in the same ASI. Classes $t3$ and $t4$ violate the ice rule and contain excited states (magnetic monopoles). Whenever a second layer is taken into account the mutual interaction between the layers must also be computed. This is accomplished in a simple way allowing a third component for $r_{ij}$ vectors, so that they also take the interaction between pairs of nanoislands belonging to distinct layers into account (clearly, vectors $\hat{e}_i$ $S_i$ are kept on layers planes, say, only with $x$ and $y$ components). If one intends to bring a third layer, one simply permits $r_{ij}$ to run over its vertices, and so on for a multilayer ASI.\\

Our simulation is carried out by considering two ASI layers parallel to each other and separated by a height offset, $h$. Each layer has $29 \times 29=841$ vertices comprising a total of $N=3480$ magnetic moments on a square lattice. [Once our system comprises a large number of dipoles and we have taken care of considering excitation in the central region of each layer, no finite size or border-like effects are expected to be appreciable, in analogy to what happens for a single layer ASI.] Our first task is to determine the ground state of the coupled layers as function of $h$. This is done by a standard Monte Carlo technique along with Metropolis algorithm implemented using Boltzmann distribution, $\sim e^{-\Delta E/k_B T}$, such that one starts off from a disordered state at high temperature; later, the system is driven to slow dynamics by cooling it to very low temperature, $\sim 0.1D/k_B$, so that the system achieves its ground state. The second task is to determine the ground state energy as a function of the separation $h$ between the layers. The energy is calculated from the Hamiltonian Eq. (1) and we have also defined the potential $V = E(h) - E_{\infty}$, where $E_{\infty}$ is the energy for decoupled layers. Third, one looks for the excitation emerging above the ground state. For this, we calculate the energy difference between an excited and the ground state, $\Delta E = E - E_0,$ as a function of the separation $h.$ We shall consider four different spin configurations: 1) a monopole pair in one layer while the another is kept in the ground state; 2) a pair of monopoles in each layer; 3) an isolated charge in one layer and another charge of the same type in the other layer; 4) opposite charges in each layer. For the latter, we also analyze the energy cost to move the charges on the layer. Finally, a basic thermodynamic analysis is performed to obtain the specific heat as a function of $h$.\\
\section{Results and Discussion}

Since $t1$ vertices bear lowest possible energy, one might think that any combination of its vertices in both layers would yield the ground state. However, this is not true at all. Actually, a single layer achieves its ground state by choosing any combination of $t1$ vertices. But, whenever mutual coupling between them is considered, then things happen in a quite different way. To better realize the scenario, we firstly remark that the ground state of a single ASI is doubly degenerate ($GS1$ and $GS2$) and populated only by $t1$ vertices. Let $t1$ class split in upper and lower vertices (see Fig. \ref{fig:fig1}); so, in both ground states, the neighbor vertices alternate $t1$-upper with $t1$-lower. In addition, if an arbitrary vertex in an ASI is $t1$-upper in the ground state $GS1$, then, the same vertex would be $t1$-lower in the ground state $GS2$. The first excited state demands the appearance of $t2$ and $t3$ vertices: $t3$ support the monopole-antimonopole pair (with unity and opposite charges $\pm 1$, blue and red spots) joined by an energetic string which is a segment of $t2$ vertices. $t4$ vertices support double-charged monopoles, demanding much higher energies to show up. Clearly, for a single layer $GS1$ or $GS2$ yields one possible ASI ground state. However, whenever taken as a combined system, BASI ground state is achieved in such a way that if one layer is at, say, $GS1$, the other must be $GS2$. In this case, the layers experience a considerable attraction, whose potential goes like $V(h)=-80(2) \, (h/a)^{-6.2(1)}$ ($V$ is measured in units of the dipolar constant, $D$). On the other hand, if the layers were put parallel exhibiting the same individual ground state, say, $GS1$ or $GS2$ in both, then they would strongly repel each other according to $V(h)=+80(2) \, (h/a)^{-6.2(1)}$ (see Fig.\ref{fig:fig1}). A similar pattern holds for $t2$ states: $t2$-up in a layer combined with $t2$-down in another yields attraction, whereas same type vertex at both layers leads to a repulsive regime. In these cases, since all vertices in $t2$ class bear net magnetization, attractive or repulsive force acquires an additional contribution due to the mutual interaction among these magnetized vertices. Another example is provided by $t1$ vertices in one layer while $t2$-class in the other, which yields to a repulsive regime. As a whole, it should be emphasized that the true ground state(s) configuration depends upon $h$; here, we have determined it for small height, $h<a$. Indeed, as $h$ becomes larger, $h\gtrsim 1.5a$, BASI is practically decoupled and one has two non-interacting ASI systems. [Indeed, its basic thermodynamics puts an even more stringent value, indicating that for $h>a$ one effectively has two decoupled layers, as discussed later].\\

Since the original degrees of freedom are dipoles interacting via the Hamiltonian (\ref{Hamiltonian}), one could suggest that we are faced with a magnetic van der Waals-like system. Before getting into such a discussion, we would like to present a brief survey of van der Waals (vdW) forces, which will help us understand the differences between both frameworks. Indeed, vdW forces arise from the mutual coupling among electric dipoles composing a system. Although two ideal dipoles interact like $r^{-3}$-potential, when effects like orientation, induction and dispersion are taken into account the net interaction must be augmented by potential functions that go like $V_{\rm vdW} (r) \sim  r^{-6}$, see Ref.\cite{London1937,Farina1999}. Such forces are keystones to understand how atoms and molecules combine to form larger clusters in gaseous, liquid, and solid substances \cite{London1937,review-Van-der-Waals}. More recently, vdW interaction has become increasingly important for a better understanding of layered compounds, including graphene and other $2{\cal D}$-materials \cite{Phys-Today-2020-vdW-2D-mat}. Indeed, in such materials the ions are held by strong covalent bonds whereas the layers experience weak out-of-plane vdW forces, allowing easy exfoliation. Even though the potential experienced by the layers presents a power-law comparable to those associated with vdW forces, the scenery is different. Indeed, in our BASI dipoles are fixed and they can only flip. They cannot rotate nor change their orientation at all, as it takes place in the Keeson picture of vdW forces. [Other effects like induction and dispersion, studied respectively by Debye and London, do not apply to our BASI at all, Refs\cite{London1937,Farina1999}]. Actually, this could be the case if ASI dipoles were free to rotate around their centers allowing orientation according to the surrounding field. This is relatively easy to be accomplished in macroscopic ASI \cite{MASI-2020}, but remains challenging to be achieved at tiny scales. There, a more feasible way to promote dipole orientation is by twisting one of the layers: although it only provides a rigid rotation of all dipoles as a whole, preliminary results have suggested that there are optimal angles at which the BASI system acquire stability \cite{workinprogress}. Perhaps, in this case we could trace back orientation effects of magnetic dipoles, making it a magnetic version of van der Waals framework.\\

For practical purpose, the repulsion between BASI layers can be exploited as a kind of magnetic levitation or magnetic damping system at nanoscale. Switching between magnetic attraction/repulsion can be achieved whenever one may drive one of the layers from $GS1$ to $GS2$. Additionally, in the realm of a bilayer system, for instance configurations $GS1$ and $GS2$ may be also viewed as being opposite density of magnetic charges, then reproducing the usual fact that opposite charges attract while like charges repel each other. By choosing those combinations carefully, one might design a kind of stable `{\em BASI rotating-saddle \cite{ThompsonCJP} configuration}', where the attractive and repulsive forces balance out. For instance, one may conceive an interesting situation where one layer is fixed while another is twisted so that attractive and repulsive forces alternate allowing oscillation to this configuration. These and other proposals are appealing nowadays since actual devices are rapidly shrinking to nanometer scale giving the possibility of tuning the magnitude of BASI magnetic force on demand.\\

\begin{figure}
	\centering
	\includegraphics[width=15 cm]{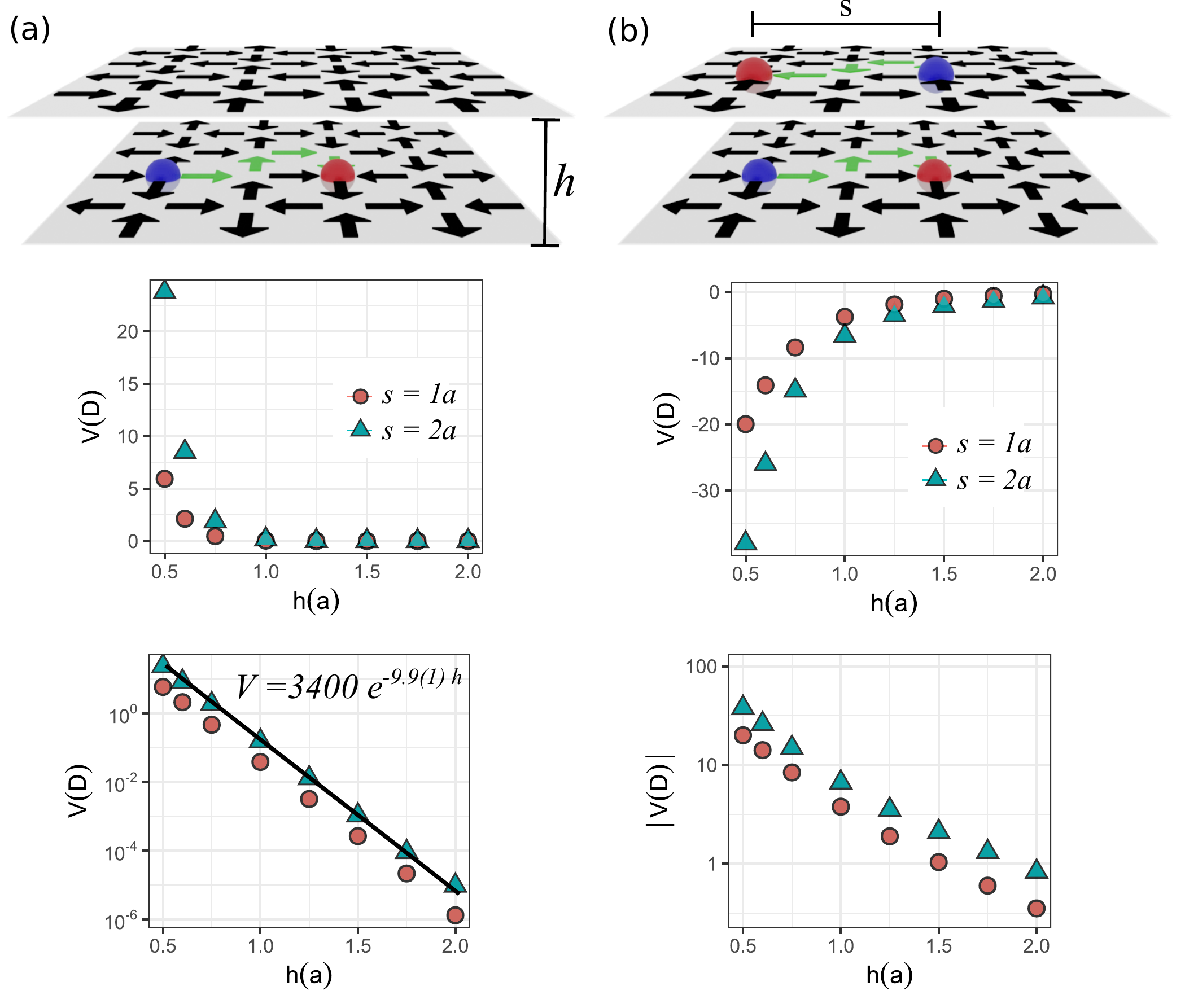}
	\caption {({\bf a}) A monopole pair lying on layer 2 (upper layer is kept at $GS1$) reinforce repulsion between the layers, which becomes strengthen as the string tension/size is enlarged. In addition, note that the interaction potential between this configuration and $GS1-GS2$ gets higher as the layers become closer. The semilog scale (bottom) shows that the potential $V$ decays exponentially as a function of $h$. The solid line shows exponential fit as $V=(3.4\pm0.2)\times10^3 e^{-(9.9\pm0.1)h}$ ({\bf b}) On the other hand, having one pair by layer with opposite poles closer, layers attract each other, as realized before with $GS1-GS2$ configuration (in this case, the points are not linearized by a semilog or log-log scale. }
	\label{fig:fig2}
\end{figure}
Now, we depart to investigate the appearance of excitations above the ground state and how they change the previous results. At each ASI layer these excitations emerge as magnetic monopole-antimonopole pairs connected by energetic strings. We then start off by considering BASI ground state given by $GS1-GS2$, say, $GS1$ in the layer 1 along with $GS2$ in the layer 2. The simplest excited state is obtained by flipping a single magnetic dipole from layer 1 (a discrete rotation of $180$ degrees; no flip is performed in layer 2, so it is kept in $GS2$). This yields a single monopole pair separated by $s=a$, with $s$ being the size of the string. Eventually, successive flips of neighbor dipoles move one of the monopoles away so that the pair is now separated by a larger (higher energetic) string, say $s=2a$, as depicted in Fig.\ref{fig:fig2}(a). The appearance of these excitations results in a repulsion between the layers: indeed, even a single monopole pair (along with the smaller string, $s=a$) is enough to overcome the attraction so that the repulsive regime dominates. Actually, whenever the string is enlarged and/or more monopoles take place, the layers repel each other with further strength. This is the fact if monopoles and strings appear in one of the layers while another is kept in its ground state. Concerning the potential itself, its exponential decay is realized and dominates at least for small string sizes, $s=a, 2a$. Further investigation is necessary to better understand such a framework and to shed light onto the physical features supporting such an emergence.\\

This leaves the question of how BASI behaves with a monopole pair in each layer. As we shall see in what follows, depending on the configuration of the monopoles, the attractive regime can be restored. Let us begin by the simplest case with a monopole pair and a string of size $s=a$, say, in layer 1 and a similar configuration in layer 2. In addition, let the poles of layer 2 be inverted with respect to those of layer 1, as depicted in Fig.\ref{fig:fig2}(b). In this case, the attraction between opposite and closest monopoles (separated by $d=h$; north/blue and south/red poles) overcomes the whole repulsion brought about by the strings and like poles interaction, so that the layers experience an attractive regime once again. If like poles are moved away their repulsion decreases considerably and attraction between layers is strengthened. On the other hand, if the configuration were like poles closer to each other, repulsion would be huge.\\
\begin{figure}
    \centering
    \includegraphics[width=15 cm]{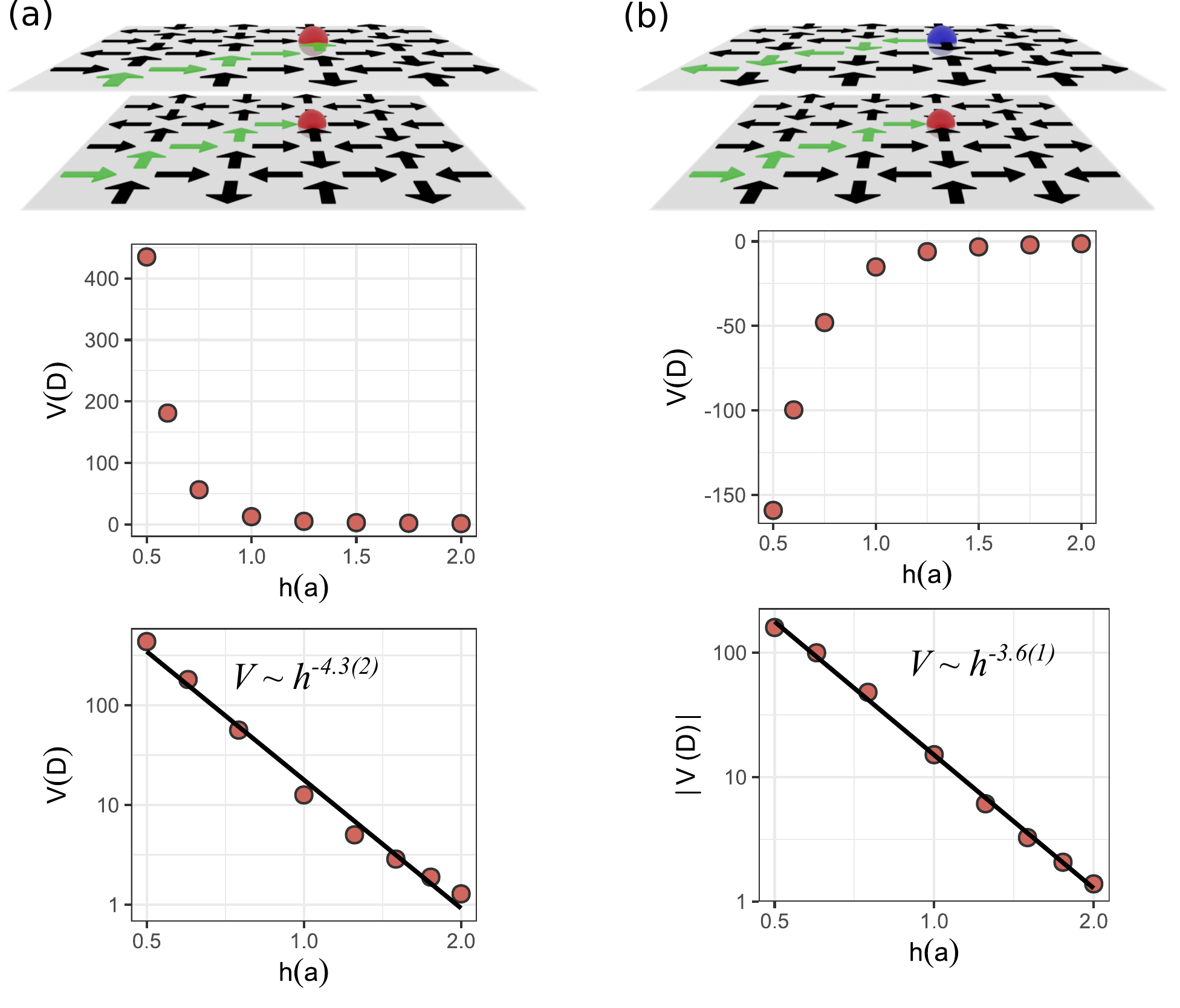}
    \caption {({\bf a}) Isolated south poles (red spots) in each layer. As expected, their interaction strengthen repulsion between layers. The solid line is a power-law fit of the form $V\sim h^\alpha$, with $\alpha=-4.3\pm0.2$ ({\bf b}) However, if opposite poles are in order, then its mutual attraction overcomes yielding attraction between the layers. The solid line is also a power-law fit of the form $V\sim h^\alpha$, with $\alpha=-3.6\pm0.1$.}
    \label{fig:fig3}
\end{figure}
Now, we would like to study the case two isolated monopoles, one placed in each layer and, whether and how the surrounding medium affects their interaction. To isolate a monopole in a layer, we should move its partner far away (which has the cost of enlarging the string until the edge of the layer, effectively expelling the moving pole outside the system). Indeed, if we consider two isolated like poles, one lying at a fixed position of each layer, as shown in Fig. \ref{fig:fig3}(a), then the layers experience a strong repulsion and tend to keep far apart. On the other hand, if one has opposite poles, their interactions overcome and the layers experience attraction once again.\\
\begin{figure}
    \centering
	\includegraphics[width=15 cm]{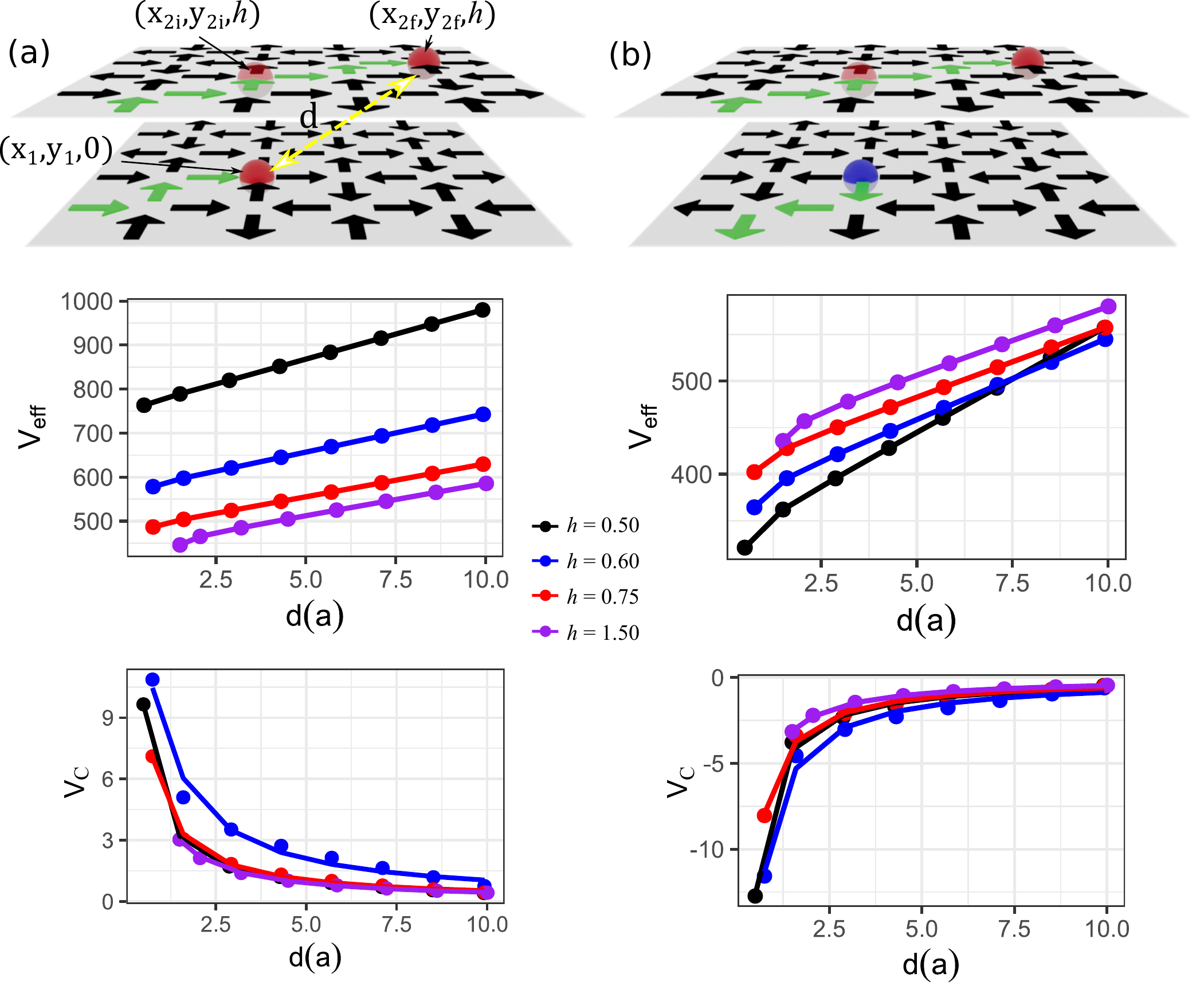}
	\caption{({\bf a}) Two like monopoles were initially separated by a vertical distance $d=h$. Then, the pole in the upper layer is displaced along $xy$ plane, as indicated.	The effective potential, $V_{\rm eff}=q^2_m\; d^{-1} + \kappa \,s +c$, against $d$ for a number of height offsets. For short monopole separation, Coulomb term dominates, while as $d$ increases the string size $s$ tends to increase even faster and the linear potential dominates. For both cases, the bottom graphics display the Coulomb potential between two magnetic monopoles, $V_C=q^2_m\; d^{-1}$, which is clearly repulsive for like charges and attractive for opposite ones.}
	\label{fig:fig4}
\end{figure}
For the sake of completeness, now the layers are kept at definite and fixed $h$ values and we intend to study the energetic of the system as one of the monopoles is displaced. For concreteness, let the monopole at bottom layer fixed at $(x_1,y_1,z_1=0)$, while the other is initially at  $(x_2=x_1,y_2=y_1,z_2=z_1+ h=h)$, but it may be displaced along $xy$ plane. As expected, also in this framework like poles repel whereas opposite monopoles attract themselves following a Coulomb potential, $d^{-1}$ ($d$ is the spatial distance between the poles, see Fig. \ref{fig:fig4}). The potential resembles that obtained for a single ASI layer, like below:
\begin{equation}\label{energetic}
	V_{\rm eff}(d,s)= \frac{q^2_m}{d}\; +\, \kappa\, s\; +\, E_c\,.
\end{equation}
where, $q_m$ is the charge of an isolated magnetic monopole (which may be positive or negative), $\kappa$ is the string tension of size $s$ which accounts for the energy cost of moving the other two monopoles far away from the remaining ones; $E_c$ is the energy cost to create these excitations, monopole pair and string as well. Table \ref{table1} presents how such parameters vary with $h$: namely, for $h= 0.5a$ one gets $q^2_m\approx 4.8 Da$ (along with $\kappa\approx 10D/a $ and $E_c\approx 450 D$). This monopole charge is comparable to that for a single square lattice ASI \cite{MolJAP2009,Mol3D,RectangularASI-1}, $\sim 3.8 Da$, but string tension and creation energy are much higher, evidencing the strong coupling between layers at this height offset. As a whole, our findings clearly show that field lines produced by magnetic monopoles lying in a layer spread radially throughout the $3{\cal D}$ space following a Coulomb-like law.\\

\begin{figure}[!h]
	\centering
	\includegraphics[width=16 cm]{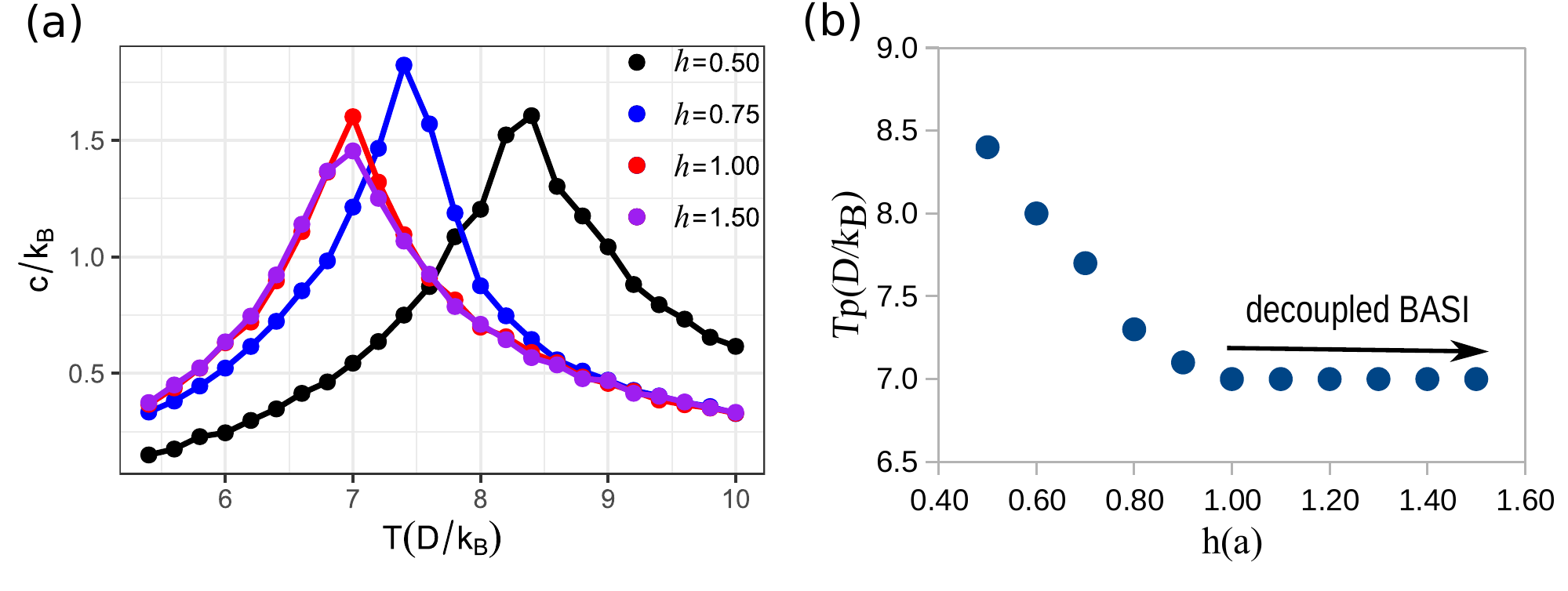}
	\caption {Simple BASI thermodynamics: ({\bf a})Specific heat, $c$, as a function of temperature, $T$. Increasing $h$ the peak is shifted to lower temperature. ({\bf b}) The peak temperature, $T_p$ against height offset, showing that BASI effectively behaves as two decoupled layers for $h>a$.}
	\label{fig:fig5}
\end{figure}

\begin{table}[!h]
	\caption{Height offset dependence of parameters $q^2_m, \kappa$ and $E_c$.} 
	\centering 
	\begin{tabular}{c c c c c c c c} 
		\hline\hline 
		& \multicolumn{3}{c}{Equal poles } &\qquad\qquad & \multicolumn{3}{c}{Opposite poles }  \\ \cmidrule{2-4} \cmidrule{6-8}
		h(a) \qquad\qquad& $q^2_m(Da)$ \qquad\qquad& $\kappa(D/a)$\qquad\qquad & $E_c(D)$ & & $q^2_m(Da)$ \qquad \qquad& $\kappa(D/a)$ \qquad\qquad & $E_c(D)$\\ [0.5ex] 
		\hline 
		0.50 \qquad\qquad& 4.8(1) \qquad\qquad& 16.13(1)\qquad\qquad & 753.7(1) & & -6.3(2)\qquad\qquad  & 16.00(3)\qquad\qquad & 333.8(3) \\ 
		0.75 \qquad\qquad& 5.3(2) \qquad\qquad& 10.67(2) \qquad\qquad& 479.7(2) & & -5.9(3) \qquad\qquad & 10.55(3) \qquad\qquad& 410.3(3) \\
		1.0 \qquad\qquad& 5.1(2)\qquad\qquad & 10.23(2) \qquad\qquad& 448.8(2)  & & -5.3(3) \qquad\qquad & 10.14(1) \qquad\qquad& 432.0(2) \\
		1.5 \qquad\qquad& 4.5(2) \qquad\qquad& 10.17(7) \qquad\qquad& 442.8(7)  & & -4.7(2) \qquad\qquad & 10.13(1)\qquad\qquad & 438.9(1) \\
		2.0\qquad\qquad & 4.2(1)\qquad\qquad & 10.16(3)\qquad\qquad & 442.2(1)  & & -4.4(1)\qquad\qquad  & 10.13(1)\qquad\qquad & 439.7(1) \\ [1ex] 
		\hline 
	\end{tabular}
	\label{table1} 
\end{table}
Finally, we deal with the basic thermodynamics of the BASI system. The specific heat, $c=\frac{\langle{E^2}\rangle - \langle{E}\rangle^2}{Nk_B T^2}$  ($k_{B}$ is the Boltzmann constant), has been obtained by standard Monte Carlo technique along with Metropolis algorithm implemented using Boltzmann distribution, $\sim e^{-\Delta E/k_B T}$, for our original array consisting by 3480 dipoles per layer. We have also implemented $10^4$ Monte Carlo steps to reach a steady state and up to $10^5$ Monte Carlo steps to obtain the averages of thermodynamic variables, each Monte Carlo step corresponding to $3480$ single-spin flips. In order to save computation time, we adopt a cutoff radius $r_c = 6a$ whenever dealing with the dipolar energy. Such a cutoff yields deviations $\le 0.1\%$ in the total energy of the system. Fig. \ref{fig:fig5} depicts specific heat as a function of temperature for distinct $h$ values. First, note that the specific heat peak is shifted to lower temperature as $h$ increases. Indeed, as the layers become decoupled the peak temperature is $T_p \approx 7 D/k_B$. This value is slightly smaller than $T=T_{c}=7.2 D/k_{B}$ reported in the works of Refs.\cite{RectangularASI-1,SilvaNJP} for a square ASI in the thermodynamics limit, whereas here we should taking into account the finite size of BASI layers.\\
\section{Conclusion and prospects}
Whenever in the ground state, the layers composing our BASI system attract each other with a force which resembles van der Waals power-law forces. When excitations emerge in the system, layers still attract if single opposite charges are present in each layer (Fig. 3b), or a monopole pair in each layer but with opposite charges closer each other. Other situations favor repulsion, making possible the realization of magnetic levitation or magnetic damping for nanoscaled systems. The switching between attraction and repulsion may be useful to design a stable {\em BASI-type rotating state} by balancing these forces. Layers twisting comes to be another ingredient to achieve other stable configurations. The dependence of specific heat peak with $h$ may be useful to determine optimal heights offset favoring stability.

As prospects, we intend to investigate how BASI behave under translation and rotation of one layer (while the other is kept fixed). As fundamental symmetries they are expected to yield novel effects and to shed further light into system physical properties.\\
\section*{Acknowledgements}
The authors thank CAPES (Finance Code 001), CNPq and
FAPEMIG for partial financial support.\\
\section*{Data Availability}
The data that support the findings of this study are available from the corresponding author upon reasonable request.\\

\end{document}